\documentclass[11pt]{article}

\usepackage{titlesec}
\makeatletter
\renewcommand\paragraph{\@startsection{paragraph}{4}{\z@}%
            {3.25ex \@plus1ex \@minus.2ex}%
            {-1em}%
            {\normalfont\normalsize\bfseries\small}} 
\makeatother

\usepackage{lineno}
\usepackage{float}
\usepackage{amsmath}
\usepackage{graphicx}
\usepackage{enumitem}
\usepackage{color}
\usepackage{amssymb}
\usepackage{authblk}

\usepackage{lineno}
\usepackage{enumitem}

\usepackage{color,soul}

\setul{0.5ex}{0.3ex}
\definecolor{Red}{rgb}{1,0.0,0.0}

\title{Freeform surface topology prediction for prescribed illumination via semi-supervised learning}

\date{}

\author[1]{Jeroen Cerpentier}
\affil[1]{KU Leuven, Department of Electrical Engineering (ESAT), Light \& Lighting Laboratory, B-9000 Gent, Belgium}
\author[1, *]{Youri Meuret}
\affil[*]{Corresponding author: youri.meuret@kuleuven.be}

\begin{document}
\maketitle
\begin{abstract}
Despite significant advances in the field of freeform optical design, there still remain various unsolved problems. One of these is the design of smooth, shallow freeform topologies, consisting of multiple convex, concave and saddle shaped regions, in order to generate a prescribed illumination pattern. Such freeform topologies are relevant in the context of glare-free illumination and thin, refractive beam shaping elements. Machine learning techniques already proved to be extremely valuable in solving complex inverse problems in optics and photonics, but their application to freeform optical design is mostly limited to imaging optics. This paper presents a rapid, standalone framework for the prediction of freeform surface topologies that generate a prescribed irradiance distribution, from a predefined light source. The framework employs a 2D convolutional neural network to model the relationship between the prescribed target irradiance and required freeform topology. This network is trained on the loss between the obtained irradiance and input irradiance, using a second network that replaces Monte-Carlo raytracing from source to target. This semi-supervised learning approach proves to be superior compared to a supervised learning approach using ground truth freeform topology/irradiance pairs; a fact that is connected to the observation that multiple freeform topologies can yield similar irradiance patterns. The resulting network is able to rapidly predict smooth freeform topologies that generate arbitrary irradiance patterns, and could serve as an inspiration for applying machine learning to other open problems in freeform illumination design.
\end{abstract}

\section{Introduction}

Optical systems to control the propagation of light play a major role in science and technology, and their importance will not diminish in the near future~\cite{caulfield2010future, capasso2018future, nikolov2021metaform}. For decades, optical engineers have relied on systems with multiple (a)spherical surfaces, which have rotational symmetry \cite{duerr2021freeform, volatier2019differential}. Recent advances in manufacturing technology however, allow the fabrication of freeform optical surfaces with completely arbitrary shape, thereby offering greater flexibility for controlling the propagation of light \cite{reimers2017freeform, zhang2021towards}. Freeform optics are widely used in imaging systems to guide the light of points in object space effectively to corresponding points in image space~\cite{jahn2017innovative, li2018two, yang2017automated, yang2015direct}. Also in the field of illumination design, freeform components are extensively used to  map the emitted light  distribution from a  source into a desired target pattern, while maintaining the luminous flux~\cite{wu2019precise, wei2020least, heemels2023limits}. The demand for such freeform illumination systems is rapidly growing, due to their application in fast-evolving fields such as optical lithography, automotive headlights and laser beam shaping~\cite{wu2018design, ibrahim2020characterization}.

Illumination optics are determined by the light source under consideration and the targeted light pattern. Their design typically comes down to calculating one or more optical surfaces that manipulate the incoming rays, in order to produce a certain prescribed irradiance distribution. Freeform illumination design methods can be separated in two categories: zero étendue algorithms and algorithms for extended light sources~\cite{wu2018design}. Zero-étendue methods are based on the assumption that the source is ideal, e.g. a point source or a collimated laser beam. Freeform design for zero-étendue sources has matured significantly, and various accurate calculation methods exist, such as the ray-mapping and Monge-Ampère methods~\cite{desnijder2019ray, wu2013freeform, madrid2022freeform, prins2014monge}. Unfortunately, the étendue of real light sources can seldom be neglected in practice. When applying zero-étendue algorithms for such extended light sources, the resulting pattern becomes blurry, and more dedicated design procedures are needed~\cite{bosel2019compact, birch2020design}. To solve this problem, wavefront tailoring and  deconvolution-based algorithms have been proposed~\cite{dross2004review,sorgato2019design, byzov2020optimization, wei2021compact}. Although these methods function well under certain specific conditions, e.g. single-chip LED emitters, they remain unsuitable for arbitrary extended light sources. Alternatively, a solution can be obtained through optimization of the parameterized freeform surface(s)~\cite{liu2012parametric, fournier2008optimization, mao2014two, situ2011combined, li2023optimizing}. These iterative methods however, are typically computationally intensive and require a significant amount of convergence time. So, as opposed to zero-étendue algorithms, the search for fast and effective freeform algorithms with extended light sources is still ongoing.

\begin{figure}[!t]
    \centering
    \includegraphics[width = 1.\linewidth]{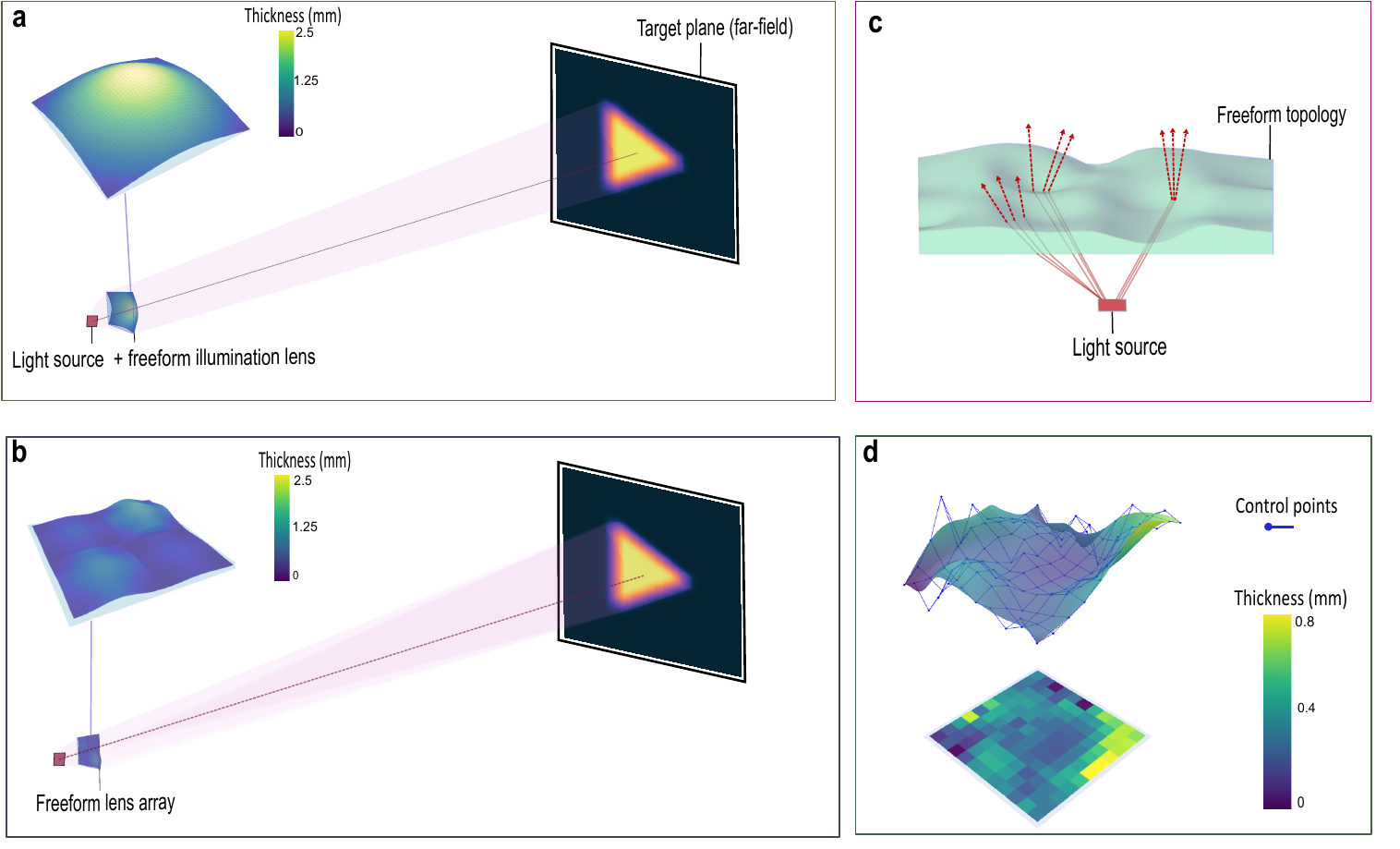}
     \caption{(\textbf{a}) Convex freeform illumination lens that generates a prescribed (far-field) target pattern from a given light source. (\textbf{b}) The light patterns generated by each of the four individual sub-lenses of a discontinuous lens array result together in the prescribed target pattern. (\textbf{c}) A smooth, continuous freeform surface \textit{topology}  that consists of multiple convex, concave and saddle-shaped regions avoids strong C2 surface discontinuities. (\textbf{d}) Such a freeform surface topology can be described as a NURBS surface that is controlled by a matrix of control points. Remark : in (a) and (b) the illumination component and illumination pattern are not drawn to scale for illustration purposes.} \label{fig:introfig}
\end{figure}

Aside from the considered light source and targeted light distribution, there is another aspect that is of practical importance in the design of freeform components for illumination purposes: the resulting freeform shape. Most freeform design methods result in overall convex or concave freeform surface shapes (Fig. \ref{fig:introfig}a). This can lead to visual discomfort glare when combined with high-brightness light sources~\cite{desnijder2018design}. A possible approach to address this important issue in lighting, is working with freeform lens arrays, where each lens element illuminates (part of) the complete target pattern. In doing so, the light flux towards each point of the target pattern is spread over the entire lens surface, similar to the case of a light diffuser. This results in a reduced brightness perception when looking towards the illumination optic, in comparison with a single-channel freeform lens~\cite{desnijder2019luminance}. Unfortunately, such freeform lens arrays have C2 discontinuities in between the individual lens elements. These discontinuities complicate manufacturing and lead to unwanted straylight~\cite{zhu2018new}. Such problems could be avoided with smooth, continuous freeform surfaces that combine multiple convex, concave and saddle shaped regions, for which we introduce the term \ul{freeform topologies} (Fig. \ref{fig:introfig}c). For collimated light sources, such an oscillating freeform topology can be used to generate broad illumination patterns with much thinner optical elements, compared with overall convex or concave freeform surfaces. Designing such smooth and shallow, oscillating freeform topologies is challenging however. Within the domain of laser beam shaping, continuous phase plates (CPP) rely on a similar surface topology, but these are mainly used for converging laser beams, and current design methods do not enable tailoring of arbitrary irradiance shapes~\cite{neauport2003design, yang2013novel}. On the other hand, such topologies could be constructed by crafting a continuous curvature between the individual convex and concave lenslets of a freeform lens array. Such a multi-stage method will likely exhibit limited generalization to create a wide range of different, complex irradiance patterns. To this day, a direct design method for freeform topologies that are capable of generating arbitrary irradiance patterns remains unpublished.

To reduce the time and complexity of optical design, researchers have recently started to use machine learning techniques. Applying deep learning to solve inverse problems in photonics and optics has only recently arisen, but the potential is indisputable~\cite{shen2017deep, wiecha2021deep, xu2021interfacing}. It may arguably become one of the main catalysts in designing complex optical configurations in the near future~\cite{gao2021deep}. In the field of computational holography for example, mature machine learning methods are already state-of-the-art~\cite{eybposh2020deepcgh, shimobaba2022deep}. Deep learning architectures for freeform design have also been presented in past research, but so far mainly within the domain of imaging optics, in order to find starting points close to a final solution~\cite{ nie2023freeform,mao2023freeformnet, chen2021generating, yang2019direct}. Within illumination design, one fully trained network has been demonstrated and the approach was restricted to creating very basic shapes~\cite{gannon2018using}. The authors of this work also underlined the need for more advanced procedures. One of the main challenges to realize a fully trained network for freeform illumination design via supervised or unsupervised learning is the lack of a fast forward operator to link the input and output parameters. Monte-Carlo (MC) raytracing is typically used to evaluate the performance of freeform optics for a given light source. Taking into account the complex shape that freeform surfaces can adopt, tens of thousands to even millions of rays are needed to model the resulting irradiance pattern with limited statistical noise.

This paper presents a deep learning framework for solving the inverse problem of finding a refractive freeform surface topology in order to obtain a prescribed irradiance distribution from a predefined light source. The framework integrates supervised learning for modeling the raytracing on one hand, and semi-supervised learning for freeform surface prediction on the other hand. This double approach alleviates the computational burden to train the freeform surface prediction fully via Monte-Carlo raytracing. We demonstrate that rapid convergence can be obtained by considering the deviation of the irradiance resulting from a predicted freeform surface with the target irradiance as a loss function during training. This approach is different from the more typical method of considering the mean average error of the predicted freeform surface shape compared to the ground truth, which proves to have inferior convergence. This behavior is linked to the observation that two or more distinct freeform topologies can result in visually identical illumination patterns. The deep learning framework rapidly provides a suitable freeform design as is illustrated for various target patterns. As such, we demonstrate that deep learning does not only serve as a tool to enhance typical optimization procedures, but that it can be used as a fast, standalone method in freeform illumination design.

\section{Methods}

\noindent\textbf{Problem statement}\indent Consider a thin refractive element, consisting of a planar entrance surface, orthogonal to the optical axis (z-axis) and a freeform surface at the exit. A smooth (C2 continuous) freeform topology can be represented as a non-uniform rational basis spline (NURBS) surface with $a \times b$  equidistant control points (Fig. \ref{fig:introfig}d). When the x- and y-coordinates of these control points are predefined, only the z-coordinate is a free parameter that samples the local height of the freeform surface. With this parameterization, the freeform component is fully characterized by an  $a \times b$ matrix ($\mathbf{\mathcal{F}}$). By restricting all z-coordinates to a certain interval [0, \textit{t}], the thickness of the refractive element can be limited. 

Light rays can be propagated from the source through the optical surfaces towards a detector surface. By binning the radiant flux of light rays on this detector, in an $m \times n$ spatial receiver grid, the irradiance distribution on the detector ($\mathcal{I}$) can be sampled. The propagation of the light through the optical system via raytracing thus corresponds with the mapping

\begin{equation}
\label{eq:1}
\{\mathbf{\mathcal{F}} \in [0, t]^{a \times b}\} \rightarrow \{\mathcal{I} \in \mathbb{R}^{m \times n}_+\}.
\end{equation}
The problem that is tackled in this work is modeling the inverse relation
\begin{equation}
\label{eq:2}
 \{\mathcal{I} \in \mathbb{R}^{m \times n}_+\} \rightarrow  \{\mathbf{\mathcal{F}} \in [0, t]^{a \times b}\}
\end{equation}
i.e. for any random irradiance  distribution $\mathcal{I}$ on the target plane, predict the $\mathcal{F}$ that results in $\mathcal{I}$ via raytracing.

\begin{figure}[!t]
    \centering
    \includegraphics[width = 1.\linewidth]{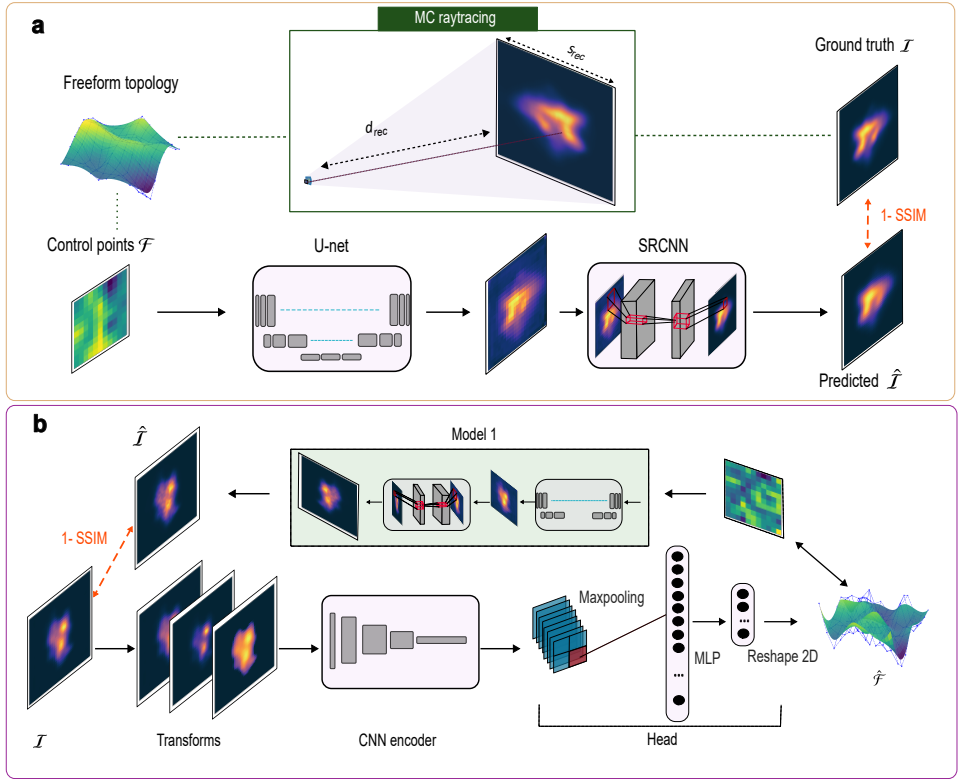}
    \caption{\textbf{(a)} Illustration of the U-net-SRCNN model for predicting the obtained irradiance on the target plane for a certain freeform surface. The model is trained by considering pairs of freeform topologies and the corresponding simulated far-field irradiance distributions, via MC raytracing. Training loss is evaluated using the SSIM loss between the MC-simulated ground truth $\mathcal{I}$ and the U-net-SRCNN-predicted $\mathcal{\hat{I}}$. \textbf{(b)} Illustration of the model to predict a freeform grid $\hat{\mathcal{F}}$ from an input irradiance $\mathcal{I}$. Training is achieved by considering the SSIM loss between the input irradiance $\mathcal{I}$ and  $\hat{\mathcal{I}}$, which is the predicted irradiance of $\hat{\mathcal{F}}$ by the first model. \label{fig:models}}
\end{figure}

\noindent\textbf{Framework structure}\indent Our deep learning framework consists of two different architectures (Fig. \ref{fig:models}a, \ref{fig:models}b). The first deep learning architecture is a U-net that models the irradiance distribution on the receiver as a result of propagating light rays through the optical system. This architecture implements the mapping relation in Eq. \ref{eq:1}  which is typically obtained with MC raytracing. U-nets are commonly used segmentation structures and are ideal for modeling 2D-to-2D mappings. They are widely used in  medical imaging and for image deconvolution~\cite{ronneberger2015u, yanny2022deep}. However, when tracing rays through arbitrary freeform surfaces, the resulting irradiance distribution often contains highly detailed features. U-nets generally predict images with relative low resolution, causing these high resolution features to  vanish during prediction. To attain more detailed irradiance  prediction, a super resolution CNN (SRCNN) is appended~\cite{dong2015image}. From a practical point of view, an input of $a \times b$ freeform control points is thus encoded into a set of feature maps, representing the local properties of the freeform surface. These local properties are then reconstructively upsampled and combined into an $m \times n$ irradiance distribution.

The full model, visualized in Fig. \ref{fig:models}a, is trained on structural similarity index measure (SSIM) loss in a supervised manner \cite{wang2004image}:
\[
\text{{SSIM}}(x, y) = \frac{{(2\mu_x\mu_y + C_1)(2\sigma_{xy} + C_2)}}{{(\mu_x^2 + \mu_y^2 + C_1)(\sigma_x^2 + \sigma_y^2 + C_2)}}
\]
where $x$ and $y$  represent the horizontal and vertical directions within an image, and $\mu$ and $\sigma$ the variance. Furthermore, $\sigma_{xy}$ is the covariance of both directions, and $C_1$ and $C_2$ are constants that stabilize the nominator and denominator in the calculation. The training data is generated using MC raytracing in the defined optical setting, producing a set of freeform-irradiance pairs ($\mathcal{F}$, $\mathcal{I}$). Moreover, data augmentation is performed by considering the inverse rotational symmetry of $\mathcal{F}$ and $\mathcal{I}$. In particular, since the freeform topology is characterized by a square grid, its rotation results in a reverse rotation of the corresponding irradiance distribution. In short:
\begin{equation}
    \label{eq:aug}
    \mbox{rot}(90, \mathcal{F}) \rightarrow \mbox{rot}(-90, \mathcal{I}).
\end{equation}
By considering this symmetry, each freeform surface can be rotated clockwise 3 times, resulting in 3 augmented freeform-irradiance pairs for training the models.

The second network architecture is designed to model the mapping relation of Eq. \ref{eq:2} and is schematically shown in Fig. \ref{fig:models}b, together with the considered learning strategy. This architecture consists of a typical CNN encoder network, containing a 2D maxpooling head and multi-layer perceptron regressor (MLP). As visualized in Fig. \ref{fig:models}b, the CNN thus encodes an irradiance distribution into a set of feature maps, of which the maximal element is chosen via maxpooling. Based on these pooled features, the freeform control points are then regressed in the MLP layer, thus predicting the surface parameters $\hat{\mathbf{\mathcal{F}}}$ for an input irradiance $\mathcal{I}$ \cite{lee2023deep}. To ensure the prediction of a freeform surface topology that yields an irradiance distribution resembling the input irradiance, it is required to capture both the local and global characteristics of the input irradiance distributions $\mathcal{I}$ in the training phase. This is achieved by using a three-channel input into the network, consisting of $\mathcal{I}$, along with its exponential and logarithmic transformations. The integration of such transforms with CNNs has proven beneficial in past research, e.g. for the enhancement of feature extraction in low-light areas within images~\cite{shen2017msr}. This is due to the increased value range in low-irradiance areas after applying a logarithmic transform, and the opposite effect follows from an exponential transform. Supplying these transforms along with the original images then allows the model to simultaneously process low-, medium- and high-irradiance areas of the input distribution.

The pretrained U-net-SRCNN model is used in the training phase of this second network for two different tasks. First of all, it is used to produce a large set of input irradiance distributions for the learning phase of the second model, without needing to run a huge amount of MC raytracing simulations. The advantage of this approach compared to using completely arbitrary irradiance distributions, is that these distributions are a result of the considered freeform topology parameterization, and can thus in principle be obtained with the considered freeform component. Secondly, the pretrained raytracing model is used during the actual training phase to produce a pseudo-labeled irradiance $\hat{\mathcal{I}}$ for any predicted $\hat{\mathbf{\mathcal{F}}}$ \cite{lee2013pseudo}. The assumption is that these generated pseudo-labels serve as legitimate irradiance distributions, reflecting the MC raytraced results. Their goal is to compare them with the input irradiance distributions during the training phase of model 2. So with $m_1$ and $m_2$ representing model 1 and model 2 respectively, the trainable loss $\mathcal{L}$ is evaluated as:

\begin{equation}
\label{eq:loss}
    \mathcal{L}\{\mathcal{I}, m_2(\mathcal{I})\} = 1 - \mbox{SSIM}\{\mathcal{I}, m_1[m_2(\mathcal{I})]\} = 1 - \mbox{SSIM}\{\mathcal{I}, m_1(\mathcal{\hat{F}})\},
\end{equation} 
with $m_1(\mathcal{\hat{F}})$ the generated pseudo-labeled irradiance to be compared with the ground truth $\mathcal{I}$. By including the (frozen) raytracing model in the training phase of the freeform prediction architecture, the emphasis of model 2 is on replicating the ground truth irradiance distribution, as opposed to replicating the  ground truth freeform surface, which is not considered in the training of the second model. 

This approach is somehow related to the unsupervised learning strategy that was adopted in computational holography, in favor of supervised learning using an extensive set of random phase masks paired with their simulated amplitude pattern~\cite{shimobaba2022deep}. A main difference is that computational holography can rely on an analytic forward/backward operator for the complex field at the image plane. For the considered case, the role of this analytic operator is taken over by the pretrained U-net-SRCNN model. 

\section{Results}

\textbf{{Optical simulation setting}}\indent The framework is applied in a specific optical setting, to illustrate its usage and performance. A planar light source of $3 \times 3$ mm square, with a lambertian radiation pattern, is illuminating a refractive element of $10 \times 10$ mm square at a distance of 40 mm, with a planar entrance surface and freeform exit surface. The refractive index of the component is 1.5, and a  maximum thickness of $t =0.8$ mm is considered. This component redirects the incident light towards a square receiver plane at a distance of $d_{rec} = 500$ mm and with a side length $s_{rec}=$ 1000 mm, i.e. in the far-field of the lens (see Fig. \ref{fig:models}a). Only rays that intersect with the refractive element are traced towards the receiver plane. The freeform surface is characterized by a matrix of $11 \times 11$ equidistant control points of the corresponding, third degree NURBS surface~\cite{piegl1996nurbs}. In order to produce irradiance patterns that cover the entire target plane, freeform surface topologies with multiple hills and valleys are necessary. This is a consequence of the shallow lens thickness and the fact that large surface slopes are needed to realize the required deflection angles. This is only possible by combining multiple positive and negative surface slopes.

The dataset for supervised learning of the U-net is generated in the commercial software LightTools~\cite{LightTools}, which allows  MC raytracing through freeform (NURBS) lens surfaces. To start, 25000 random freeform topologies are generated by selecting an arbitrary z-coordinate within the chosen interval for each lens surface control point. The corresponding irradiance distribution $\mathcal{I}$ for each freeform topology $\mathcal{F}$ is then calculated by tracing 15000 rays from the light source towards the receiver plane with a spatial receiver grid of 50 $\times$ 50 bins. To reduce roughness in distributions that are spread out across the entire receiver surface area, a small smoothing kernel of 3 $\times$ 3 is implemented. Training is executed on the resulting $\{\mathcal{F}, \mathcal{I}\}$ pairs, using a 90-10 train-validation split. More details on the architectures and training procedure in this specific setting are added in Appendix A.

\begin{figure}[!t]
    \centering
    \includegraphics[width=1.\linewidth]{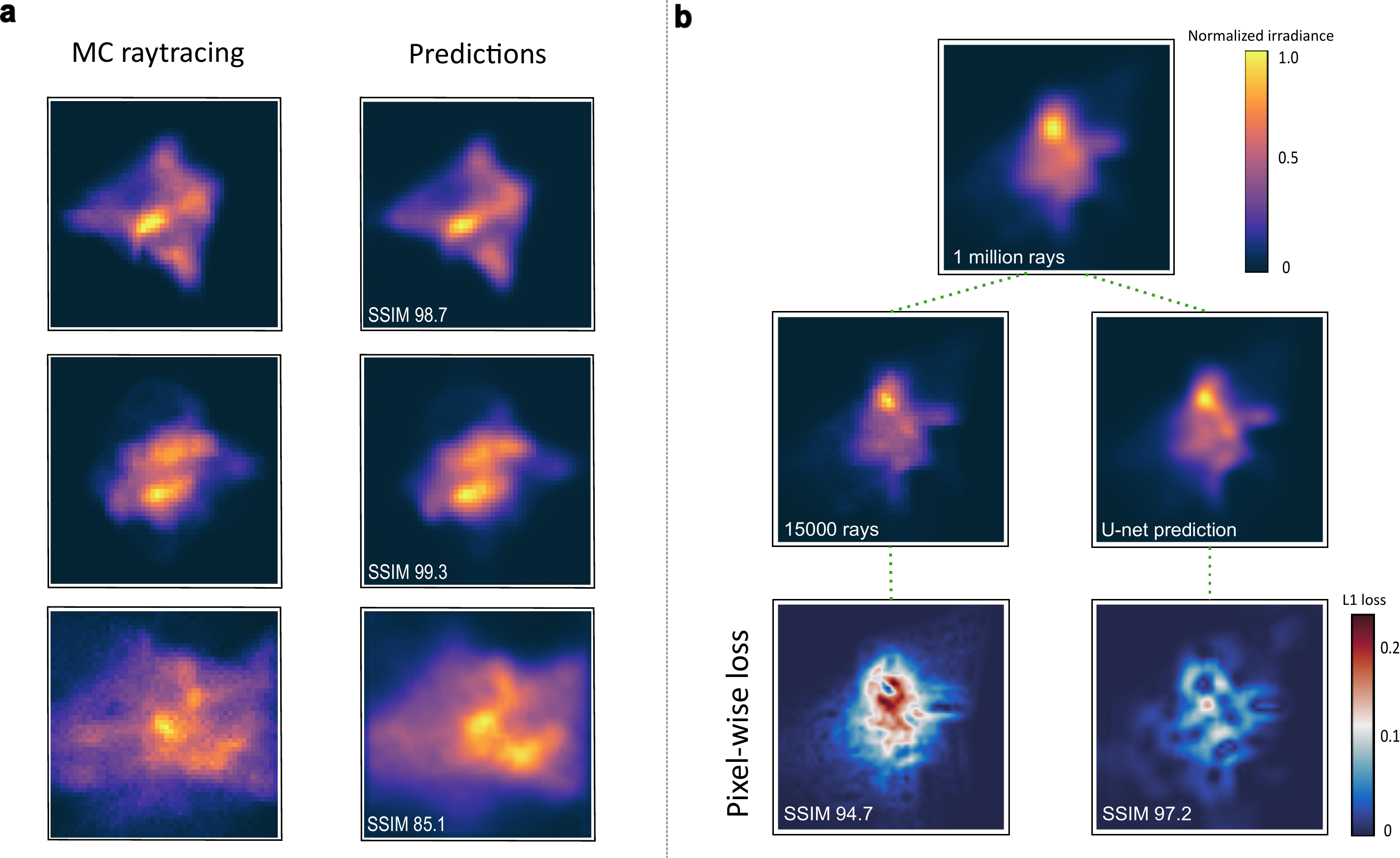}
     \caption{\textbf{(a)} MC raytracing simulated irradiance versus predicted irradiance by the U-net-SRCNN model. The bottom image represents a low SSIM outlier in the validation set. \textbf{(b)} Comparison of pixel-wise errors: the irradiance simulated using 1 million rays is compared with the originally simulated irradiance using 15000 rays (left), and the U-net prediction, as shown on the right. Additionally, SSIM values are calculated for each comparison.\\ Since the focus of this study lies in irradiance shapes, each irradiance was individually normalized.} \label{fig:unetresults}
\end{figure} 

\noindent \textbf{{Irradiance prediction}}\indent Fig. \ref{fig:unetresults}a shows quantitative and visual results for the U-net-SRCNN model. The predicted irradiance distribution by the model is compared with the corresponding Monte-Carlo raytracing result for three different cases, together with the SSIM loss. The bottom figure shows  an outlier in terms of SSIM, compared to the average SSIM of the complete validation set, which is 98.3\%. The average 1.7\% error is likely due to the simulation noise in the considered raytracing simulations, since 15000 MC rays are traced towards a 50 $\times$ 50 receiver grid; following a typical $1/\sqrt{N}$ approximation with $N=15000/50/50$, the noise per bin is estimated at 41\%. A more extensive simulated rayset would be required to suppress these noise effects, but this turned out to be unnecessary. Indeed, in Fig. \ref{fig:unetresults}b, the validation irradiance for a random SSIM outlier is compared with the MC simulated irradiance for the same freeform lens but with a rayset of  1 million rays. Taking the absolute pixel-wise deviation of the predicted irradiance by the U-net with the newly raytraced irradiance as a measure, and by comparing it with the pixel-wise deviation between the original and newly raytraced samples, the differences are smaller when comparing the  predicted irradiance with the simulated sample with 1 million rays, even though the model has been trained on the original, noisy samples. This thus demonstrates that the model is not solely capable of reproducing raytracing, but it also denoises the samples that it was trained on. It is important to note that in recent research, U-net CNNs have already been suggested to be useful for fast irradiance evaluation of freeform components~\cite{tang2023fast}. However, this previous architecture considered only 49 control points with an inference time of 67 ms, which is around 6 times slower than our proposed architecture. For the proposed semi-supervised learning approach, inference speed of this U-net is crucial to ensure efficient training. Furthermore, the raysets for MC raytracing were also large ($2 \times 10^6$), leading to a time-consuming data generation process.

Given that the proposed model is capable of generating accurate irradiance distributions, it can be used as a rapid alternative for the MC raytracing simulations. Following this logic, a synthetic dataset of 3.4 million freeform $\mathcal{F}$ - irradiance $\mathcal{I}$ pairs was generated within minutes. This enables learning on a large synthetic dataset of input irradiance distributions that are a result of the considered freeform topology, enhancing generalization and reducing overfitting for the more complex, reverse freeform prediction task. The 3.4 million irradiance samples were again separated in a 90-10 train-validation split.

\begin{figure}[!t]
    \centering
    \includegraphics[width=1.\linewidth]{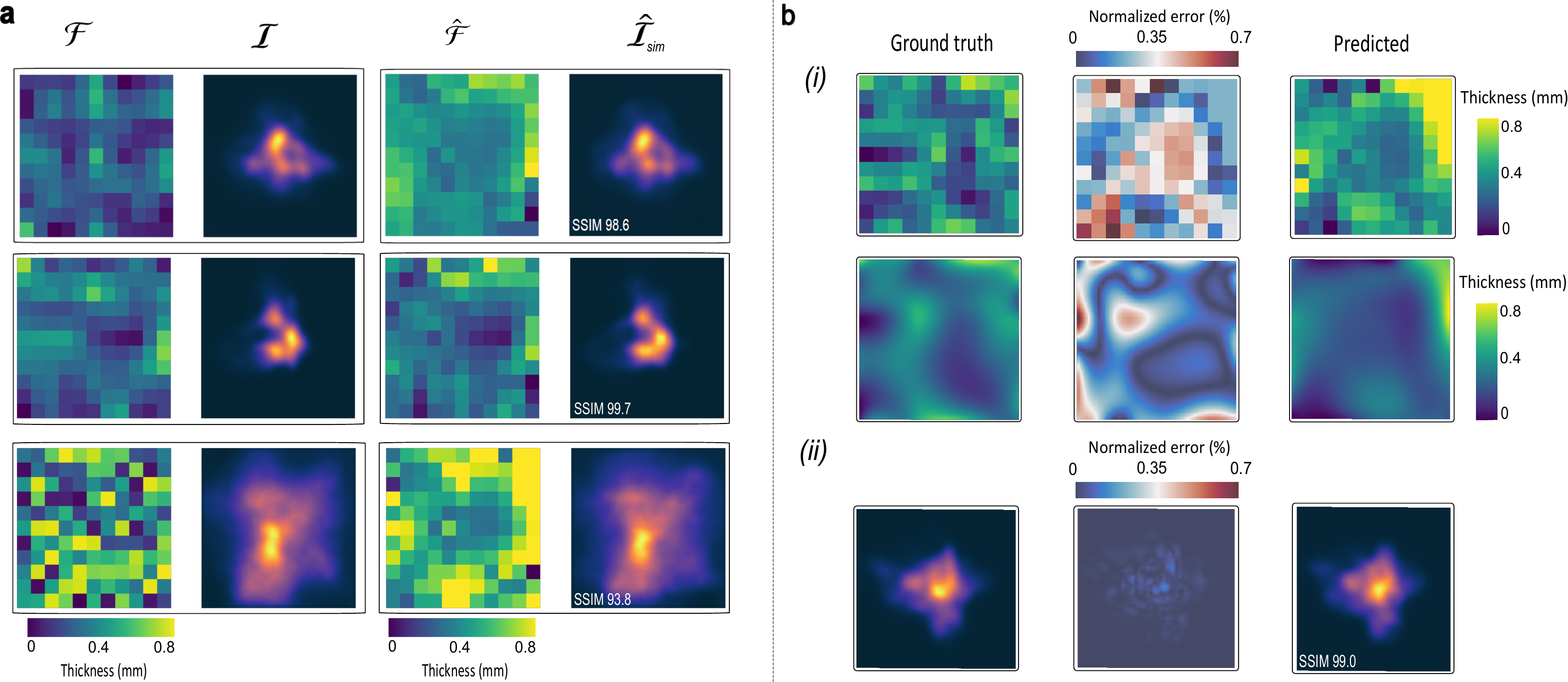}
    \caption{\textbf{(a)} Freeform prediction results for three validation cases. Ground truth ($\mathcal{F}, \mathcal{I}$) against the predicted freeform matrix ($\mathcal{\hat{F}}$) and corresponding MC simulated irradiance ($\hat{\mathcal{I}}_{sim}$) using 1 million rays. \textbf{(b)} (\textit{i}) Pixel-wise error for the normalized predicted freeform surface control points $|\mathcal{F} - \mathcal{\hat{F}}|$ as well as the interpolated NURBS surfaces. (\textit{ii}) Pixel-wise error for the simulated irradiances. Notice the significant difference in freeform surface topology (\textit{i}) while producing an almost identical irradiance distribution (\textit{ii}).\label{fig:ffmodelresults}}
\end{figure}

\noindent \textbf{Freeform prediction on validation data.} Fig. \ref{fig:ffmodelresults}a shows quantitative and visual results for the freeform prediction model. Inference is done in 11 ms for a single sample, and can be further enhanced with a GPU-specific inference optimizer, such as NVIDIA's TensorRT module. The predicted results are studied for three validation cases, where in each case, a ground truth freeform $\mathcal{F}$ - irradiance $\mathcal{I}$ pair is compared with the predicted freeform control points $\mathcal{\hat{F}}$ by the CNN encoder network and the resulting irradiance distribution with this predicted freeform ($\hat{\mathcal{I}}_{sim}$). The resulting irradiance distribution has been simulated with MC raytracing using an extensive rayset of 1 million rays, rather than with the U-net model used in the training. This assures a fair assessment of the capabilities of the model to predict an $\mathcal{\hat{F}}$ that results in a prescribed $\mathcal{I}$ via raytracing. A quantitative assessment of the freeform prediction accuracy is provided  by again considering the SSIM between the input $\mathcal{I}$ and the raytraced irradiance distribution $\hat{\mathcal{I}}_{sim}$. The visual similarity between $\mathcal{I}$ and $\hat{\mathcal{I}}_{sim}$ illustrates the performance of the developed framework. However, one also notices the visual discrepancy between $\mathcal{F}$ and $\mathcal{\hat{F}}$. Fig. \ref{fig:ffmodelresults}b  considers this discrepancy more in detail for one specific case, and shows that the pixel-wise deviation for the freeform control points and the interpolated NURBS topologies is much higher than for the irradiance distributions, a feature that is witnessed for most of  the validation cases. This observation supports the hypothesis that radically different freeform surface topologies can produce visually identical irradiance distributions. Therefore, training for SSIM on the pseudo-labeled U-net irradiance versus the targeted irradiance is a more logical approach than training for mean average error of the predicted freeform surface compared to the ground truth.

\begin{figure}[!t]
    \centering
    \includegraphics[width=1.\linewidth]{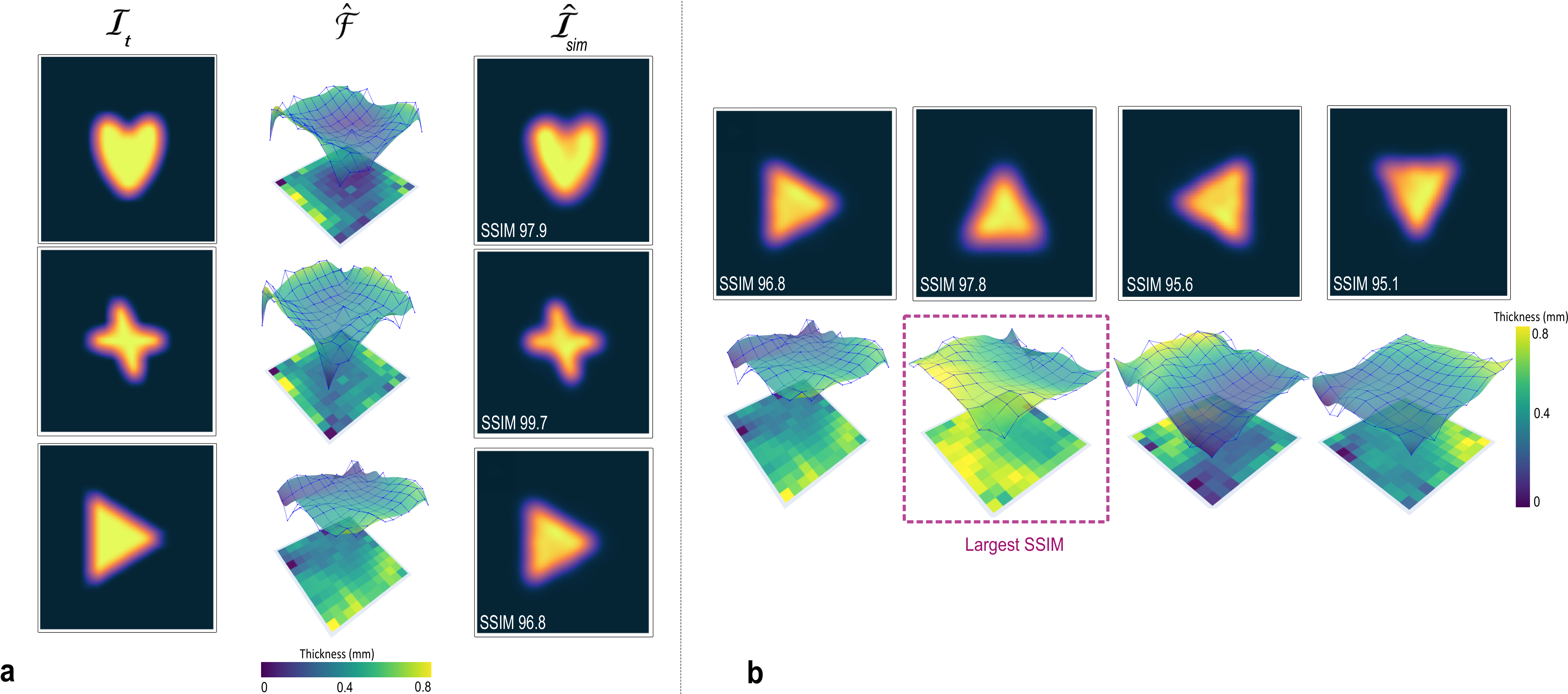}
     \caption{Freeform prediction results for certain  custom target irradiances $\mathcal{I}_t$. \textbf{(a)} $\mathcal{I}_t$ against the predicted freeform surface topologies $\hat{\mathcal{F}}$ and the MC raytraced irradiance $\hat{\mathcal{I}}_{sim}$, resulting from $\hat{\mathcal{F}}$. \textbf{(b)} Rotate test-time augmentation on the triangular target. The results illustrate the existence of various   freeform surface topologies that produce almost visually identical irradiance distributions. \label{fig:predictionresults}}
\end{figure} 

\noindent \textbf{Freeform prediction on custom targets.} The model performance is also verified in terms of predicting freeform topologies that produce prescribed irradiance distributions, which are neither in the training nor in the validation set. This is the main target of the developed framework. Fig. \ref{fig:predictionresults}a shows an overview of the results for three chosen irradiance patterns ($\mathcal{I}_t$). The predicted freeform surface topology ($\hat{\mathcal{F}}$) and corresponding raytraced irradiance pattern ($\hat{\mathcal{I}}_{sim}$) are given, together with the SSIM of the simulated pattern with respect to the prescribed irradiance. The results indicate that the model is capable of producing fairly complicated freeform surface topologies that match the target irradiance distributions. Still, one could wonder if the model produces the best freeform topology prediction in terms of SSIM, since there is no raytraced ground truth in this case. As a test, a \textit{rotate test-time augmentation} is applied  by considering all 90$^{\circ}$-rotations of the target pattern. The obtained freeform topologies with the resulting MC-simulated irradiance and corresponding SSIM values are shown in Fig.~\ref{fig:predictionresults}b. These results re-affirm that multiple freeform topologies can result in nearly-identical irradiance patterns and the model finds one of these topologies. In other words, the inverse problem is ill-posed, at least within the limitations of MC raytracing. From a practical point-of-view however, such rotational test-time augmentations, or similar alternatives, could prove interesting to generate multiple solutions, out of which the best performing, most smooth or most oscillating option could be selected, depending on the specific application.

\begin{figure}[!t]
    \centering
    \includegraphics[width=.85\linewidth]{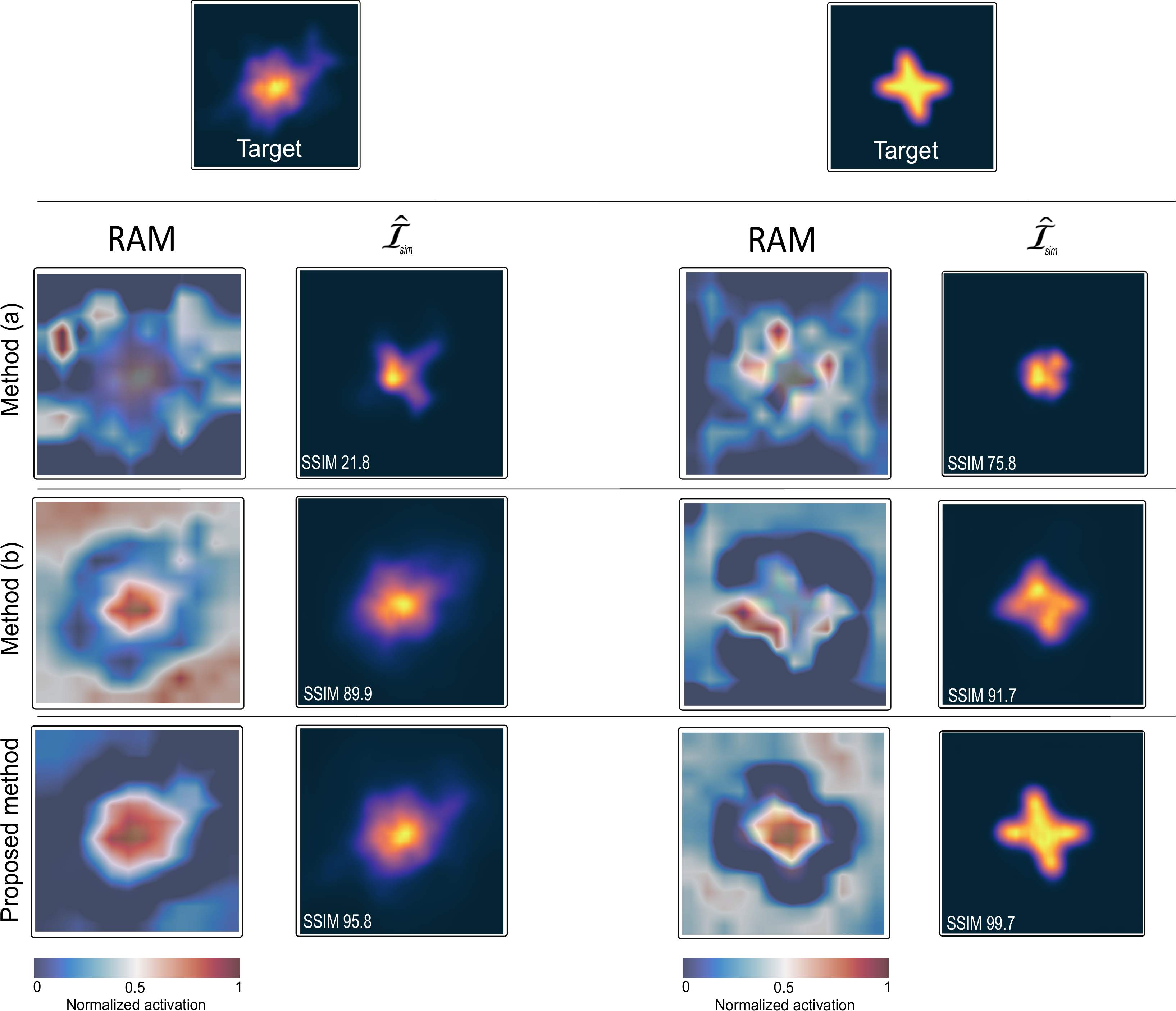}
    \caption{Performance comparison of three different learning approaches. Regression activation maps (RAM, normalized) and the produced irradiance ($\hat{\mathcal{I}}_{sim}$) of the predicted freeform topology $\mathcal{\hat{I}}$ are shown, both for a target irradiance from the validation set and a custom target irradiance. \label{fig:train-results}}
    
\end{figure}
\noindent \textbf{Training performance.} 
Finally, the efficiency of the proposed training approach is assessed through a comparative analysis with two alternative methodologies.

\begin{enumerate}[label=(\alph*)]
    \item Training in a supervised manner, on L1 loss between $\mathcal{F}$ and $\mathcal{\hat{F}}$, using the full synthetic dataset.
    \item Training on the SSIM loss with pseudo-labeled irradiances,  but using the 25000 MC raytraced irradiance samples instead of the predicted irradiances by the U-net-SRCNN model. 
\end{enumerate}

Fig. \ref{fig:train-results} considers two target distributions to illustrate the main performance differences: an irradiance distribution from the validation set linked to an actual freeform topology, and a custom prescribed irradiance. For both cases, the regression activation maps (RAM) and resulting  raytraced irradiances $\hat{\mathcal{I}}_{sim}$ from the predicted freeform topologies are shown for method (a) and  (b), compared to the  proposed approach. Regression activation maps are included since they provide a detailed representation of how a model localizes discriminative regions affecting the regression outcome~\cite{wang2017diabetic}. Comparing RAM for the first target pattern, it is visible that method (a)  fails to produce confident feature maps at relevant locations. In comparison, model (b) manages to locate the core of the supplied irradiance as the most relevant area, but some noise remains. The proposed training method clearly delivers the most accurate localization, with activation maps that overlap with the target distribution. This results in the highest SSIM value for the corresponding raytraced irradiance. Similar results can be seen for the custom target distribution. In this case, the benefit of relying on synthetic data over the base MC raytraced data (method b) is visually clear when looking at the obtained irradiance distribution.

\begin{figure}[!t]
    \centering
    \includegraphics[width=.95\linewidth]{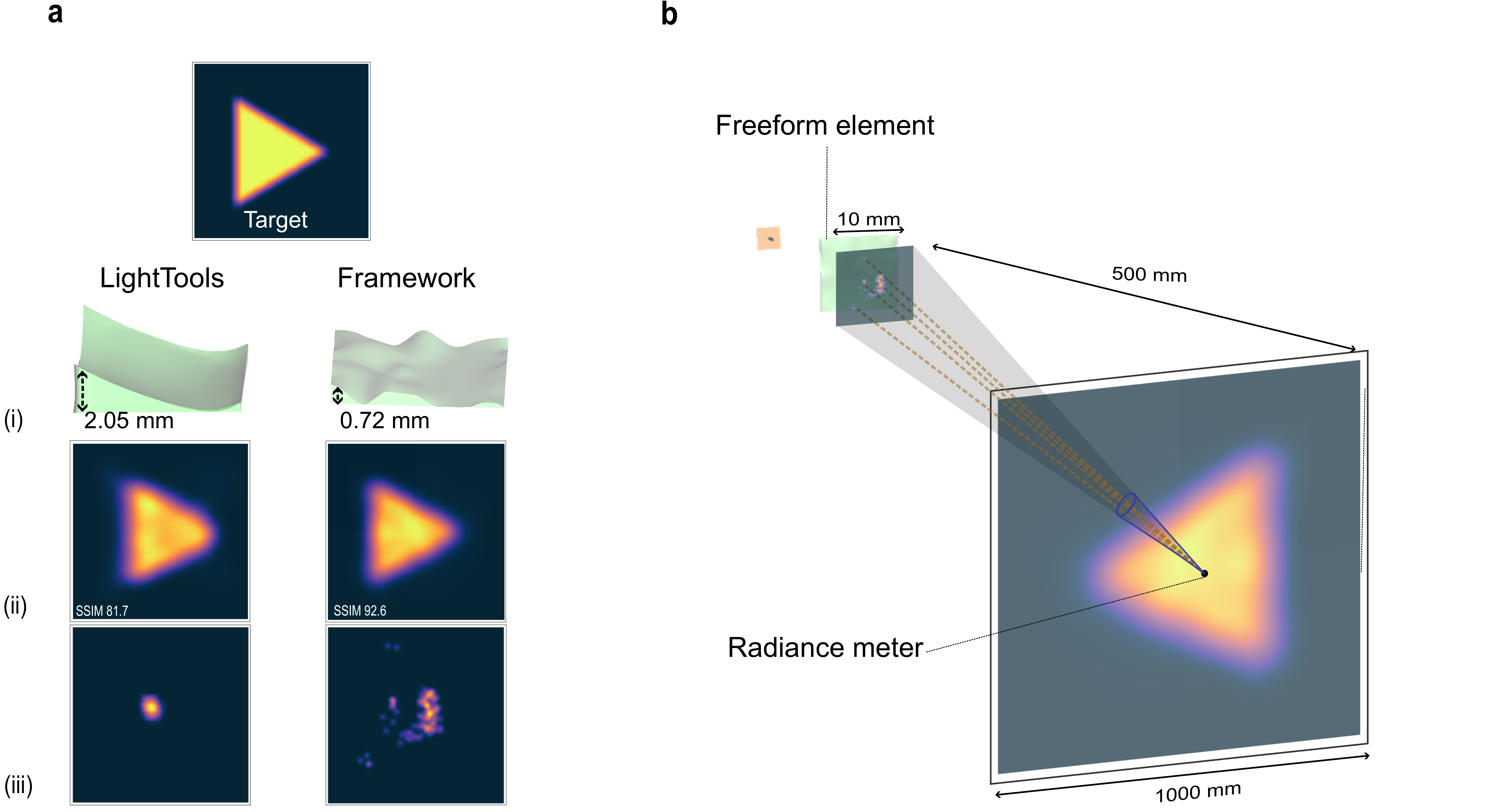}
    \caption{(a) Comparison between freeform surfaces generated by LightTools and the proposed framework, in terms of (i) freeform surface shape, (ii) resulting irradiance distribution at receiver plane and (iii) radiance distribution at freeform component. (b) The spatial spreading of the radiance distribution at the illumination component, towards a specific receiver direction, can be sampled with a radiance meter in the LightTools software. The size of the freeform element was not drawn at scale for illustration purposes.} \label{fig:glarefig}
    
\end{figure}

\paragraph{Comparison with state-of-the-art}

A goal of this study was to produce freeform surface topologies that differ from classical freeform surfaces with an overall convex/concave shape.   To illustrate this capacity, a generated freeform topology is compared  with the solution that is calculated by the LightTools Freeform Design Feature tool, using also $11 \times 11$ control points. In both cases, a large uniform triangle is considered as target irradiance pattern. Fig. \ref{fig:glarefig}a shows the comparison between the resulting NURBS surfaces in both cases. Notice the expected concave shape in the LightTools generated freeform surface versus the varying freeform topology generated by our framework. This results in a 2.8$\times$ smaller surface height. The resulting irradiance distributions are shown in (ii). In both cases, the limited amount of control points results in clear visual deviations from the target pattern;  however, the SSIM of the solution that is generated by LightTools is approx. 10\% lower than the solution by the  proposed method. The oscillating shape of the topology might raise concerns about TIR losses; however, with fresnel losses enabled in the raytracing simulation, the transmission efficiency is 91.7\%, being only 0.6\% lower compared to the efficiency of the classical freeform surface (92.3\%).

Finally, it is verified that the freeform surface topology, results in a more spread out light distribution from the exit surface towards each point of the target pattern. To do this, both freeform components are simulated in LightTools, and a small radiance meter is positioned at the receiver surface, with its field-of-view aimed at the freeform exit surface.  In this way, the radiance at the exit from the illumination component is sampled from a specific receiver direction. In the case of the concave lens, light is only propagating from a specific position/area on the freeform surface, towards a corresponding position in the target pattern. There is thus a one-to-one mapping of the freeform surface to the target pattern. In the case of the freeform surface topology however, light is propagating from multiple positions/areas on the freeform surface, towards each position in the target pattern, as is illustrated in (iii). This spatial spreading of the radiance distribution over the entire illumination component is an effective approach to avoid visual discomfort glare.

\section{Discussion}

This paper presents a semi-supervised learning framework for predicting refractive freeform topologies that produce a certain target irradiance. In contrast to prior work on machine learning  for freeform design that predominantly relies on 1D multi-layer perceptron-like networks with contextual information, this study employs 2D convolutional neural networks (CNN) to model the relationship between the obtained irradiance and freeform topology. To train the network, a U-net is used  for modeling the Monte-Carlo raytracing of light from a predefined light source, in order to generate pseudo-labeled irradiance distributions that are compared with the input target irradiances. We demonstrate that this semi-supervised learning approach for freeform topology prediction is superior compared to a supervised learning approach using ground truth freeform topology/irradiance pairs. The resulting framework offers an end-to-end solution for rapid, smooth freeform topology design to produce an arbitrary prescribed target irradiance. The framework is trained within a specific optical setting, using a restricted parameterization for the freeform lens topology. This implies that the model is only capable of predicting freeform components within this parameterization and for the considered illumination setting. Despite these limitations, the proposed framework offers significant opportunities regarding the design and implementation of novel freeform optics.

The freeform components that are introduced in this paper, with their smooth oscillating surface topology and resulting limited thickness, could serve as a novel beam shaping technology. Current freeform micro-optical components always consist of a periodic array of multiple individual lenses or mirrors~\cite{aderneuer2021surface, bec2019broadband}, and while the fabrication of such components has evolved significantly over the past years, the inevitable grooves in between the different lens/mirror elements still pose significant manufacturing challenges in order to avoid stray light~\cite{ kumar2022advances}. The smooth surface topologies that result from the proposed network avoid such strong surface C2 discontinuities. The resulting beam shaping components can be used to generate non-paraxial target distributions, when illuminated with (nearly) planar incident wavefronts from LEDs or lasers.

The proposed framework and training method could be extended to other, more general freeform illumination design problems. A straightforward extension would be a surface topology parameterization with more control points. This requires up-scaling of the model, for which larger CNN variants, or even transformers, could be used. Along with the extension of the model, more data will also be required as well as upscaling of the irradiance resolution. A larger set of control points could allow the generation of even more complex irradiance targets, as well as the representation and design of larger/thinner optical components with more convex/concave/saddle-shaped regions. However, it should be clear that the level of detail is also limited by the size of the light source in comparison to the distance to the freeform component, when only one freeform surface is considered. A more interesting expansion from a practical point of view, could therefore be the extension of the framework to arbitrary (extended) light sources and multiple freeform surfaces. This may be realized by using the spatial distribution of the light source as an additional input. Predicting multiple freeform surfaces could be achieved by increasing the final MLP output head size. Of course, this would also require a significant increase in the amount of training data. For the implementation of a flexible machine learning model for general freeform illumination, it is clear that computation power will be a must.

While the presented semi-supervised learning strategy proves superior to a supervised learning strategy, for the prediction of shallow freeform surface topologies, it remains to be seen if this would also be the most effective strategy for the design of overall convex-concave freeform surfaces. The training of such a framework would certainly require a more specific surface parameterization that enforces concaveness or convexness. Alternatively, contextual data about the freeform shape could be added in the training phase of the 2D-CNN, e.g. by concatenating an encoded prompt about the surface shape in the final linear layer of the network. Whatever the outcome, also in this case, the use of an additional network for modeling the raytracing will result in much faster and more effective training.           

The discussion above makes clear that the investigation of advanced machine learning techniques for the design of freeform illumination optics has only been started. In this respect, the proposed framework can serve as a demonstration that deep learning allows rapid and standalone freeform optical design.

\appendix

\section*{Appendix A: Framework specifications}

\paragraph{Raytracing model (U-net)}

\begin{figure}[!t]
    \centering
    \includegraphics[width = .95\linewidth]{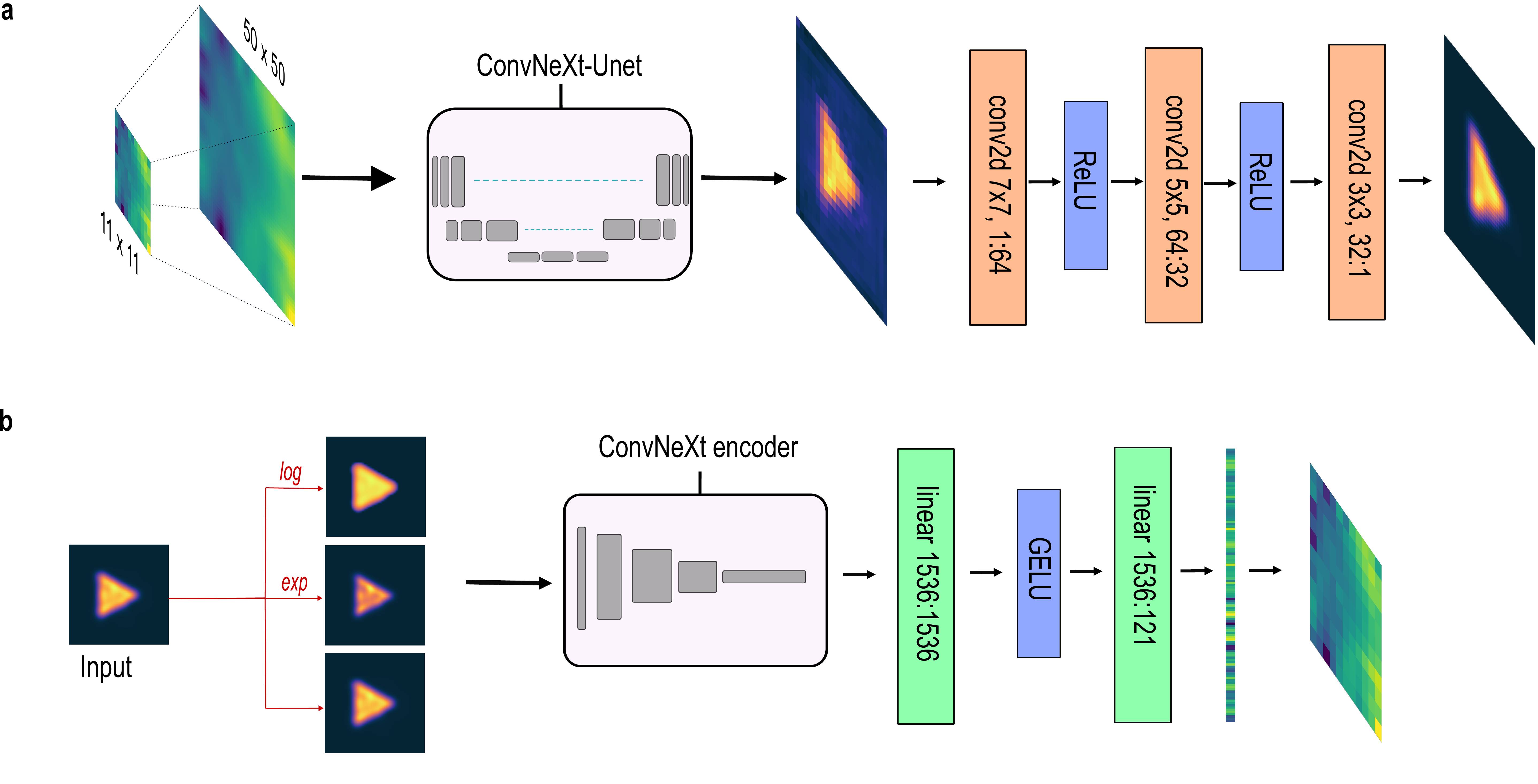}
     \caption{(\textbf{a}) U-net-SRCNN model with visualization of how the control points are interpolated and how the SRCNN constructs the final irradiance distribution. \textbf{(b)} The freeform prediction model, with highlights on the input transformations and the MLP structure that predicts the freeform surface control points.} \label{fig:methodsfig}
\end{figure}

The raytracing model receives an 11 $\times$ 11 matrix of control points that unambiguously determine a freeform surface topology. This matrix is interpolated to match the convolution kernels (bilinear, 50 $\times$ 50). The resulting matrix then passes through a U-net encoder-decoder structure, with the encoder being the state-of-the-art ConvNeXt~\cite{liu2022convnet} (Fig. \ref{fig:methodsfig}a). These feature maps are then decoded into a corresponding illuminance distribution via convolutions. This U-net output is a 25 $\times$ 25 pixel irradiance, which is upscaled to the irradiance resolution of 50 $\times$ 50 pixels via bilinear interpolation. Consequently, the upscaled images are passed through a trainable SRCNN architecture~\cite{dong2015image} with kernel sizes of 7, 5 and 3, respectively. Such SRCNN uses non-linear, e.g. ReLU, activations and convolutions to enhance image resolution by filling in some of the high-frequency image details.

\paragraph*{Freeform prediction model}

The main component of the freeform prediction architecture is a CNN encoder, which again is ConvNeXt~\cite{liu2022convnet}. The model takes a 3-channel image as input, consisting of:
\begin{equation}
    \{\mathcal{I}, \mbox{log}(\mathcal{I}), \mbox{exp}(\mathcal{I})\}.
\end{equation}
Using the logarithmic and exponential transforms of $\mathcal{I}$ allows the model to explore the global and local properties more easily (Fig. \ref{fig:methodsfig}b). The model head contains a 2D maxpooling layer, which is a $2 \times 2$ filter that runs over all the extracted feature maps, maintaining the maximal value as the output. This significantly reduces overfitting. The maxpooling output is passed through a non-linear mapping multi-layer perceptron that outputs 121 variables, with GELU activation. The output is then reshaped into a $11 \times 11$ grid $\hat{\mathcal{F}}$ of freeform topology control points. 

\paragraph*{Training setup}

All models in the proposed method use ImageNet-21k pre-trained weights~\cite{ridnik2021imagenet}, although the contribution of these is likely limited. Training is done using exponential learning rate decay with linear warm-up. Training and testing was performed on an NVIDIA RTX 3070 GPU with 8GB VRAM. A full overview of the training, as well as the used encoder variant, is shown in Table \ref{table:training}. For the freeform prediction models, the training procedure is given for each of the methods discussed in the training performance section.

       \begin{table}[!ht]
\centering

\resizebox{.975\textwidth}{!}{%

\begin{tabular}{c|c|ccc}
     &\textbf{U-net-SRCNN} & &\textbf{ Freeform prediction}&\\ 
     & & Method (a) & Method (b) & Proposed method \\
    \hline
    Optimizer&  AdamW\cite{loshchilov2017decoupled} &  AdamW\cite{loshchilov2017decoupled}&  AdamW\cite{loshchilov2017decoupled}&  AdamW\cite{loshchilov2017decoupled}\\
    Base learning rate& 1e-4 & 6e-4& 2e-4& 6e-4\\
    Weight decay& 0.05& 0.05& 0.05& 0.05\\
    Warmup epochs& 3, linear& 5, linear& 1, linear& 5, linear\\
    Learning rate scheduler& cosine decay & cosine decay & cosine decay & cosine decay\\
    Training epochs& 50& 20& 25& 20\\
    Batch size & 512 & 196 & 512 & 196 \\
    Augmentations & randomflip$[$Eq. \ref{eq:aug}$]$ & randomflip$[$Eq. \ref{eq:aug}$]$ & randomflip$[$Eq. \ref{eq:aug}$]$ & randomflip$[$Eq. \ref{eq:aug}$]$ \\
    Encoder variant & \textit{ConvNeXt-base}~\cite{liu2022convnet} & \textit{ConvNeXt-large}~\cite{liu2022convnet} & \textit{ConvNeXt-tiny}~\cite{liu2022convnet} & \textit{ConvNeXt-large}~\cite{liu2022convnet} \\
    Drop path rate & 0.05 & 0.1 & 0 & 0.1 \\
    Full data size & 25000 & 3.4 million & 25000 & 3.4 million

\end{tabular}}

\caption{Full training set-up for all the considered models. \label{table:training}}
\end{table}


\paragraph*{Acknowledgements}
\noindent Agentschap Innoveren en Ondernemen (HBC.2020.2713).

\paragraph*{Data availability}

\noindent Data underlying the results presented in this paper are not publicly available at this time but may be obtained from the authors upon reasonable request.

\bibliographystyle{naturemag}
\bibliography{sample}
\end{document}